\documentclass[12pt]{iopart}

\expandafter\let\csname equation*\endcsname\relax
\expandafter\let\csname endequation*\endcsname\relax

\usepackage{amsmath}
\usepackage{graphicx} 
\begin{document}

\title{Theory of Josephson current on a lattice model of grain boundary in $d$-wave  superconductors}

\author{Takashi Sakamori$^1$, Satoshi Kashiwaya$^1$,
Rikizo Yano$^1$,Yukio Tanaka$^1$,Takafumi Hatano$^2$ and Keiji Yada$^1$}

\address{$^1$ Department of Applied Physics, Nagoya University, Nagoya 464-8603, Japan}
\address{$^2$ Department of Materials Physics, Nagoya University, Nagoya 464-8603, Japan}
\ead{sakamori.takashi.e7@s.mail.nagoya-u.ac.jp}
\vspace{10pt}
\begin{indented}
\item[]March 2023
\end{indented}

\begin{abstract}
    Identifying the origins of suppression of the critical current at grain boundaries of high-critical-temperature superconductors, such as cuprates and iron-based superconductors, is a crucial issue to be solved for future applications with polycrystalline materials.
    Although the dominant factor of current suppression might arise during material fabrication and/or processing, investigating it due to an internal phase change of the pair potential is an important issue in understanding the threshold of the critical current.
    In this paper, we study the Josephson current on a symmetric [001]-tilt grain boundary (GB) of a $d$-wave superconductor on a lattice model.
    In addition to the suppression of the maximum Josephson current associated with the internal phase change of the $d$-wave pair potential which has been predicted in continuum models, we find a unique phase interference effect due to folding of the Fermi surface in the lattice model. 
    In particular, the resultant maximum Josephson current at low-tilting-angle regions tends to be suppressed more than that in preexisting theories.
    Because similar suppressions of the critical current at GBs have been reported in several experimental works, the present model can serve as a guide to clarify the complicated transport mechanism in GBs.
\end{abstract}

%
\vspace{2pc}

%
\submitto{\SUST}
%
%
%

\section{Introduction}
The application of high-critical-temperature superconductors (HTSCs), such as cuprates and iron-based superconductors (IBSs), to superconducting cables and joints has been the subject of research.
It requires realizing a stable persistent current even if the circuit includes the grain boundaries (GBs) of the crystals because these boundaries inevitably exist inside cables and at their joints.
Despite the successful development of superconducting cables based on low-critical-temperature (low-$T_c$) superconductors, the realization of high-current-density cables based on HTSCs requires additional techniques for crystalline orientation control, such as ion-beam-assisted deposition.
The critical current flow through GBs has been experimentally confirmed to be seriously suppressed when the tilting angle of the GB exceeds approximately 5$^\circ$ \cite{Held,Dimos,Hilgenkamp,Ivanov} for cuprates and 10$^\circ$ for IBSs \cite{Durrell,katase,Si,Iida2019}.
Two origins of this severe suppression of the critical current have been proposed.
One proposed origin is related to material issues, such as crystal growth, doping inhomogeneities, and impurity scattering around GBs \cite{Browning,Mannhart,Schmehl}.
The other proposed origin is related to the anisotropic pairing states of superconductors.
Unlike conventional low-$T_c$ superconductors, the pair potentials of cuprates and IBSs exhibit phase changes in the momentum space.
According to the current--phase relation (CPR) of the Josephson current, the internal phase change contributes to the reversal of the current direction.
Although the dominant origin of the suppression might be the former in actual cables and joints, the influences of the latter should also be clarified in detail.\cite{hammerl}
\par
Focusing on cuprates, we note that the Josephson current formula for $d$-wave superconductor junctions has been developed mainly on the basis of the continuum theory.
Sigrist and Rice \cite{SigristRice} investigated the influence of the anisotropic pair potential on the Josephson current and proposed a theoretical reversal of the Josephson current direction due to the internal phase change. 
A reversal of the current's direction accompanies the transition of a 0-junction to a $\pi$-junction.
This formula successfully explains the anomalous magnetic field dependence of superconducting quantum interference devices and the spontaneous magnetic flux in tri-crystals \cite{Wollman1,Harlingen,Tsuei1994,Tsuei2000}.
These experimental results have been accepted as strong evidence for $d$-wave pairing symmetry of cuprates. 

Another critical effect expected for $d$-wave Josephson junctions is the formation of zero-energy Andreev bound states (ZEABSs) at the interface \cite{Hu94,TK95,Kashiwaya00,ABSR2}. 
The presence of ZEABSs is predicted to enhance the higher harmonics of the CPR of the Josephson current \cite{TKJosephson,TKJosephson2,YBJosephson}. 
The nonmonotonic temperature dependence of the critical current \cite{Testa,Tafuri_2005}, the nonsinusoidal CPR \cite{Ilichev1}, and the higher harmonics of the CPR \cite{Ilichev2} due to the ZEABS have been detected experimentally in GB junctions of cuprates, supporting the validity of the theoretical models.
The character of the CPR determines the free-energy minimum because the CPR is given by the derivative of the free energy with respect to the phase difference $\phi$.
When the free-energy minimum locates at $\phi=0$ ($\pi$), it is referred to as a 0 ($\pi$)-junction.
In the presence of higher harmonics, the free-energy minimum might appear at neither $\phi=0$ nor $\pi$ but at $\phi=\phi_m$; this junction is classified as a $\phi$-junction\cite{TKJosephson,TKJosephson2} 
\par
The dependence of the maximum Josephson current on the tilting angle in HTSC GBs is an unresolved issue.
The experimentally detected suppression of the critical current at GBs \cite{Dimos} is more drastic than the suppression predicted by the continuum models.
The influences of the short coherence length have been introduced in the calculation using lattice models \cite{Tanuma98,Tanuma99}. 
Although the nonmonotonic temperature dependence of the critical current and the existence of the $\varphi$-junction was reproduced in a lattice model \cite{Shirai}, the tilting-angle dependence was inconsistent with the experimental observations.
These facts indicate that the suppression mechanism needs to be clarified beyond the mechanisms treated in the preexisting models to explain the Josephson current through the GBs.

\par
In the present work, we study symmetric [001]-tilt GB Josephson junctions using a lattice model in which the sites at the interface are shared by the lattices on the left and right sides (see figure \ref{fig_lattice-structure}).
In this model, the periodicity of the unit cell $m$ in the direction parallel to the interface changes in accordance with the tilting angle and the corresponding Brillouin zone (BZ) parallel to the interface is folded.
When the tilting angle is reduced, folding of the BZ becomes significant with increasing $m$. 
This effect strongly enhances the interference of the internal phase change of the anisotropic pair potential.
We found that, as a result of the enhanced interference, the Josephson current is substantially suppressed with increasing $m$.

The remainder of this paper is organized as follows.
In section 2, we explain the model and formulation. 
In section 3, the results of numerical calculations are presented. 
In section 4, we summarize the obtained results and clarify the physics in the proposed model.

\section{Model and formulation}
\begin{figure}[htbp]
\centering
\includegraphics[width=.5\columnwidth]{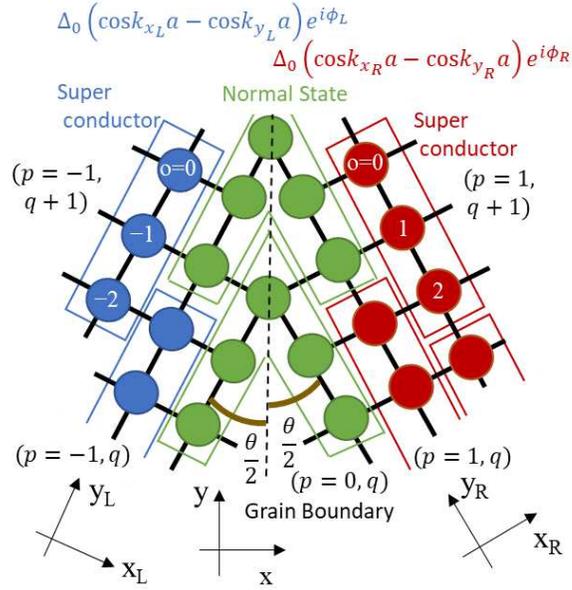}
\caption{Lattice structure of the (210) GB. Blue, red, and green solid circles denote the lattice points of the left superconductor, right superconductor, and the normal metal site, respectively. The $x_{L,(R)}, y_{L(R)}$ axes are in a tilted geometry.}
\label{fig_lattice-structure}
\end{figure}

We consider symmetric [001]-tilt GBs as shown in figure \ref{fig_lattice-structure}, where ($m10$) and ($\bar{m}10$) surfaces are connected by sharing the most end sites. 
Hereafter, this structure is referred to as the ($m10$) GB. 
The ($m10$) and ($\bar{m}10$) structures are tilted by $\pm\theta/2$ from the untilted square lattice with coordinates $(ia, ja)$ at the lattice points, where $i$ and $j$ are integers and $a$ is the lattice constant.
The coordinate of the lattice point $x<0$ is $ia\left(\cos\frac{\theta}{2}, \sin\frac{\theta}{2}\right)
 +ja\left(-\sin\frac{\theta}{2}, \cos\frac{\theta}{2}\right)$;
for $x>0$, it is 
$ia\left(\cos\frac{\theta}{2}, -\sin\frac{\theta}{2}\right)
 +ja\left(\sin\frac{\theta}{2}, \cos\frac{\theta}{2}\right)$.
We consider the ($m10$) structure for $x>0$.
As shown in figure \ref{fig_lattice-structure}, the ($m10$) structure has discrete translational symmetry along the $y$ direction by $\sqrt{m^2+1}a$. 
That is, the $x$ coordinate of the lattice point at $i=i_0$ and $j=j_0$ is equivalent that at $i=i_0-p$ and $j=j_0-mq$ with integers $p$ and $q$ that index the unit cell. 
Thus, we consider the unit cell composed of $m$ lined lattice points along the $y_{R(L)}$ direction. Two translation vectors $\hat{u}_{R(L)}$ and $\hat{v}$ are then given by
\begin{eqnarray}
\label{eq._transition-vector}
\hat{u}_{R(L)}&=\left( a\left(\sin\frac{\theta}{2}+\cos\frac{\theta}{2}\right), (-) a\left(\sin\frac{\theta}{2}-\cos\frac{\theta}{2}\right)\right)\nonumber\\
\hat{v}&=\left(0,a\sqrt{m^2+1}\right).
\end{eqnarray}

We label the lattice point by three integers $o$, $p$, and $q$; its coordinate $r(o,p,q)$ is 
\begin{align}
\label{eq._transition-opq}
\hat{r}(o,p,q)=
o\left(\sin\frac{\theta}{2}, -\mathrm{sgn}(p)\cos\frac{\theta}{2}\right)+p\hat{u}+q\hat{v},
\end{align}
where $o$ is the sublattice index, which takes values from 0 to $m$ for $p\ge 1$, and from 0 to $-m$ for $p\le -1$. 
For $p=0$, $o$ takes values from $-m$ to $m$. 
The corresponding lattice sites for the (210) GB are shown in figure \ref{fig_lattice-structure}. 
To consider a superconductor/normal metal/superconductor Josephson junction, we assume that the left (blue) and right (red) sites are in the superconducting states with the phases of $\phi_L$ and $\phi_R$, respectively, and that the center (green) sites with $p=0$ are in the normal state.
The phase difference between the left and right $\phi$ is given by $\phi_L-\phi_R$.
For $p=0$, the ($m10$) structure at $x<0$ and the ($\bar{m}10$) structure at $x>0$ share the same lattice point with $o=0$. 
The resultant unit cell of the normal state has a dog-legged shape with $2m+1$ lattice points.

For this lattice structure, we consider the tight-binding model.
The original bulk Hamiltonian without tilting is a single-band model with a $d$-wave pair potential given by
\begin{align}
\mathcal{H}_{R(L)}=&\sum_{k}(\varepsilon_k-\mu) c_k^\dag c_k+\sum_k \Delta_k c_k^\dag c_{-k}^\dag+h.c.,\\
\varepsilon_k=&-2t(\cos k_x+\cos k_y),\\
\Delta_k^{R(L)}=&\Delta_0(\cos k_x-\cos k_y)e^{i\phi_{R(L)}},
\end{align}
where $t$ and $\Delta_0$ are the hopping integral and the pair potential, respectively.
The matrix components of the Hamiltonian in Bogoliubov--de Gennes form in real space for $x \ge 0$ and $1\le o\le m$ are then
\begin{eqnarray}
\langle o,p,q|\mathcal{H}|o-1,p+1,q\rangle&=&
\begin{pmatrix}
-t & \Delta(p)\\
\Delta^*(p) & t
\end{pmatrix},\\
\langle o-1,p,q|\mathcal{H}|o,p,q\rangle&=&
\begin{pmatrix}
-t & -\Delta(p)\\
-\Delta^*(p) & t
\end{pmatrix},
\end{eqnarray}
and 
\begin{eqnarray}
\langle o=0,p,q|\mathcal{H}|o'=m,p,q+1\rangle&=&
\begin{pmatrix}
-t & \Delta(p)\\
\Delta^*(p) & t
\end{pmatrix},\\
\langle o=m,p,q|\mathcal{H}|o'=0,p+1,q-1\rangle&=&
\begin{pmatrix}
-t & -\Delta(p)\\
-\Delta^*(p) & t
\end{pmatrix}.
\end{eqnarray}
Similarly, we can consider corresponding matrix components for the ($\bar{m}10$) structure in $x<0$. 
Here, because the layer with $p=0$ is in the normal state,
\begin{align}
\Delta(p)=\left\{
\begin{matrix}
e^{i\phi_R} \Delta_0 /2 &(p\geq1)\\
0 &(p=0)\\
e^{i\phi_L} \Delta_0 /2 &(p\leq1)\\
\end{matrix},\right.
\end{align}
Using these matrices, we calculate the Green's function at the center sites and obtain the Josephson current through the normal-state sites.

To calculate the Josephson current, we define the current operator $J$ from the equation of continuity in the lattice model:
\begin{align}
    e\frac{d}{dt}\{c^{\dagger}_{ky,0,0}(t)c_{ky,0,0}(t)\}=J_{LC}(k_y)-J_{CR}(k_y).\label{eq:continuity}
\end{align}
where $c_{ky,p,o}$ ($c_{ky,p,o}^\dag$) is the annihilation (creation) operator at the site of sublattice $o$ in the $p$th layer with momentum $k_y$. 
The left-hand side of Eq. (\ref{eq:continuity}) is obtained by the Heisenberg equation, and we obtain the current operator $J_{CR}(k_y)$:
\begin{align}
    J_{CR}(k_y) = \frac{iet}{\hbar}(c^{\dagger}_{ky,0,1}c_{ky,0,0}-c^{\dagger}_{ky,0,0}c_{ky,0,1}+e^{ik_y}c^{\dagger}_{ky,0,m}c_{ky,0,0}-e^{-ik_y}c^\dagger_{ky,0,0}c_{ky,0,m}),
\end{align}
The expectation value of $J$ is calculated by the Green's function at $\tau=0$, where $\tau$ is the imaginary time. 
Because we consider the stationary state, $J_{ky}(\phi)=\langle J_{LC}(k_y) \rangle=\langle J_{CR}(k_y)\rangle$ and $\langle J_{CR}(k_y)\rangle$ is given by
\begin{align}
    \langle J_{CR}(k_y) \rangle = \frac{iet}{\hbar}(G_{C,0,1}(\tau=0,k_y)-G_{C,1,0}(\tau=0,k_y)\nonumber\\+e^{ik_y}G_{C,0,m}(\tau=0,k_y)-e^{-ik_y}G_{C,m,0}(\tau=0,k_y)).
\end{align}
The Josephson current through the normal sites $J(\phi)$ is given by
\begin{align}
    J(\phi)=\frac{1}{\sqrt{m^2+1}a}\int J_{ky}(k_y,\phi)dk_y,
\end{align}
 because the sites at $x=0$ are aligned in the $y$ direction with a period of $\sqrt{m^2+1}a$.

The Green's function at $\tau=0$ is obtained by the Matsubara Green's function:
\begin{align}
    G_C(\tau=0,k_y)=\frac{1}{k_BT}\sum_{n}G_{C}(i\epsilon_n,k_y),
\end{align}
where $\epsilon_n$ is the Matsubara frequency.
To calculate the Matsubara Green's function in this junction system, we first calculate the functions in the semi-infinite system: $G_L(i\epsilon_n)$ for $p\le -1$ and $G_R(i\epsilon_n)$ for $p\ge1$.
The details of the calculation of $G_L(i\epsilon_n)$ and $G_R(i\epsilon_n)$ are given in the appendix.
The Green's function at $p=0$ in the ($m$10) GB is then given by
\begin{align}
    G_{C}(i\epsilon_n) = (i\epsilon_n I-H_{0,0}-H_{0,1}G_R(i\epsilon_n) H_{1,0}-H_{0,-1}G_L(i\epsilon_n) H_{-1,0})^{-1},
\end{align}
where $H_{i,j}$ is the matrix $H_{i,j}=\langle p=i|\mathcal{H}|p=j\rangle$.

We also calculate the Josephson current for the continuum model using the following equations 
for $d$-wave superconductor junctions \cite{TKJosephson,TKJosephson2}:
\begin{align}
    R_NJ = \frac{\pi\overline{R_N}k_BT}{e}\sum_{\epsilon_n}\int_{-\pi/2}^{\pi/2}F(\varphi,i\epsilon_n,\phi)\sin(\phi)\sigma_N\cos(\varphi)d\varphi,
\end{align}
\begin{align}
    \epsilon_n=2\pi k_BT\left( n+1/2 \right),
\end{align}
\begin{align}
    F(\varphi,i\epsilon_n,\phi)=\frac{2\Delta(\varphi_+)\Delta(\varphi_-)}{\Omega_{n,+}\Omega_{n,-}+\epsilon^2_n+\{1-2\sigma_N\sin^2(\phi/2)\}\Delta(\varphi_+)\Delta(\varphi_-)},
\end{align}
\begin{align}
\Omega_{n,\pm}= \rm{sgn}(\epsilon_n)\sqrt{\Delta^2(\varphi_{\pm})+\epsilon^2_n},
\end{align}
\begin{align}
    \Delta(\varphi_{\pm})=\Delta\cos(2(\varphi\mp\theta/2))e^{i\phi},
\end{align}
\begin{align}
    \sigma_N=\frac{\cos^2(\varphi)}{\cos^2(\varphi)+Z^2},
\end{align}
\begin{align}
    \overline{R_N}^{-1}=\int_{-\pi/2}^{\pi/2}\sigma_N\cos(\varphi)d\varphi,
\end{align}
and $Z$ is the barrier parameter.

The corresponding formula for $s$-wave superconductor junctions is \cite{Furusaki91}
\begin{align}
    R_NJ=\frac{\pi\Delta_0}{2e}\frac{1}{\sqrt{1-\sigma_N\sin^2(\phi/2)}}\tanh\left(\frac{\Delta_0}{2k_BT}\sqrt{1-\sigma_N\sin^2(\phi/2)}\right)\sin\phi.
\end{align}

\section{Results}
We used $\mu = -t, k_BT_c=10^{-3}t, T=0.05T_c$; $\Delta_0$ was taken to be Bardeen–Cooper–Schrieffer (BCS)-like in the calculations given by
\begin{align}
    \label{eq._BSC}
    \Delta_0(T)=\Delta_0(T=0)\tanh\left( 1.74\sqrt{\frac{T_c-T}{T}}\right).
\end{align}
\begin{align}
    \Delta_0(T=0)=1.76 k_BT_c.
\end{align}
The maximum value of $\Delta_{k}^{L,R}$ is 1.5 $\Delta_0$.
\begin{figure}[htbp]
\centering
\includegraphics[width=.4\columnwidth]{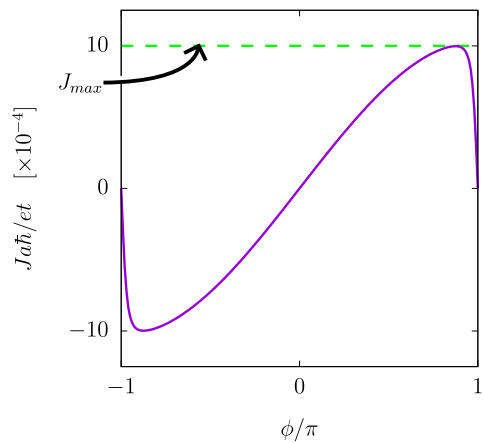}
\caption{Current-phase relation (CPR) for a (100) grain boundary (GB). The horizontal axis shows the phase difference and the vertical axis shows the Josephson current.}
\label{fig_CPR_0}
\end{figure}
\begin{figure}[htbp]
\centering
\includegraphics[width=.4\columnwidth]{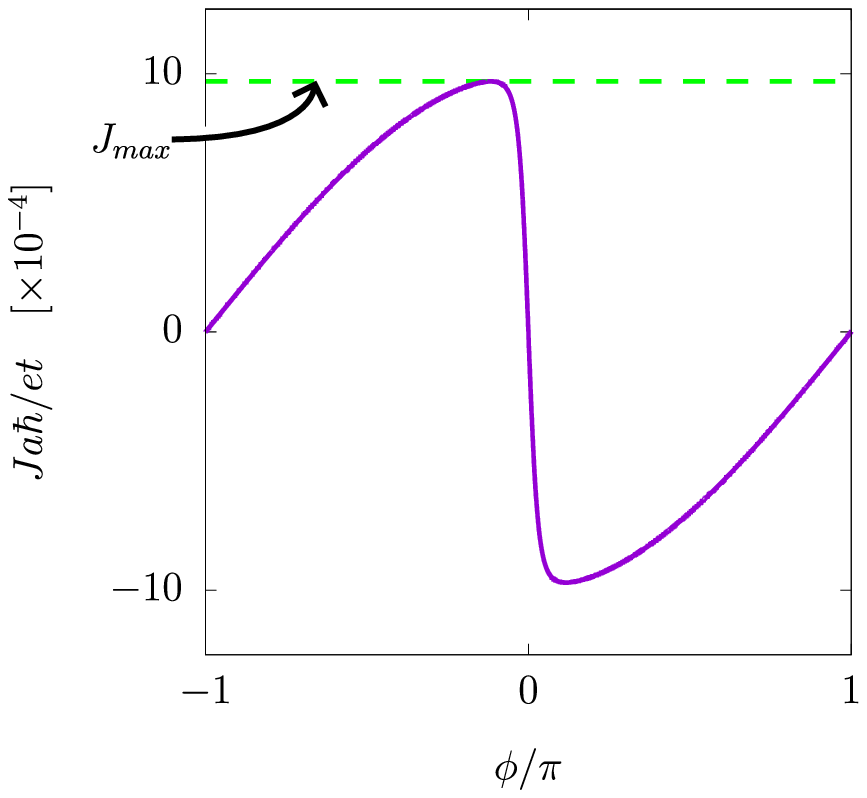}
\caption{CPR for the (110) GB. The horizontal axis shows the phase difference, and the vertical axis shows the Josephson current.}
\label{fig_CPR_pi}
\end{figure}
\begin{figure}[htbp]
\centering
\includegraphics[width=.4\columnwidth]{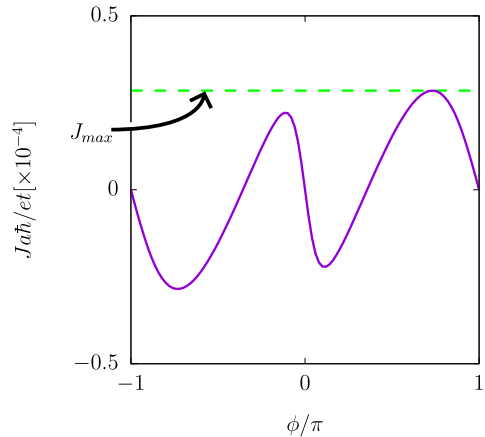}
\caption{CPR for the (210) GB. The horizontal axis shows the phase difference, and the vertical axis shows the Josephson current.}
\label{fig_CPR_phi}
\end{figure}

\par
By changing $\phi$, we obtained the CPR (figures \ref{fig_CPR_0} to \ref{fig_CPR_phi}).
Because the present superconducting state does not break the time-reversal symmetry, the CPR is expanded as
\begin{equation}
    J(\phi)=\sum J_n \sin(n\phi).
\end{equation}
The CPR for a (100) GB (figure \ref{fig_CPR_0}) exhibits standard sinusoidal 0-junction behavior with $J_1>0$, whereas it exhibits $\pi$-junction behavior with $J_1<0$ for the (110) GB (figure \ref{fig_CPR_pi}).
Because $J_1$ changes sign, a region exists where $|J_1|$ becomes smaller than the other higher harmonics.
In this case, the CPR shows so-called $\phi$-junction behavior \cite{TKJosephson,TKJosephson2}, where the higher harmonics of the CPR are enhanced (figure \ref{fig_CPR_phi}). 
The maximum Josephson current $J_{max}$ is obtained from the maximum value of $J(\phi)$.
We then calculate $J_{max}$ for ($m10$) GBs for $m=1$ to 17. We also calculate $J_{max}$ for the case of $m=\infty$ (i.e., the (100) GB). Among these structures, the (110) and (100) GBs have a perfect crystalline square lattice structure without defects.
\par
\begin{figure}[htbp]
\centering
\includegraphics[width=.4\columnwidth]{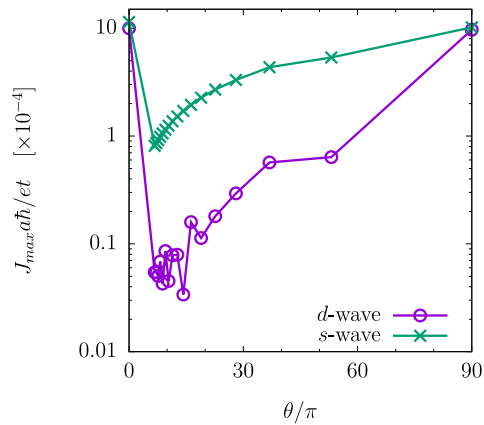}
\caption{Maximum Josephson current $J_{max}$ for $s$-wave (cross) and $d$-wave (circle) junctions with tilting angle $\theta$.}
\label{fig_Jc_theta}
\end{figure}
Figure \ref{fig_Jc_theta} shows the $J_{max}$ of ($m10$) GBs for the $s$-wave and $d$-wave cases. 
First, $J_{max}$ for the $s$-wave pair potential decreases with decreasing tilting angle $\theta$, except in the case of $\theta=0^\circ$, even though the pair potential does not exhibit the phase change. 
This suppression of $J_{max}$ depends on the character of the present lattice model. 
In the present model, ($m10$) GBs are connected by sharing the sites at $x=0$ with periodicity $\sqrt{m^2+1}a$ in the $y$ direction. 
The number of conducting channels per length along the $y$ direction is then proportional to $1/\sqrt{m^2+1}=1/\sqrt{\tan^{-2}(\theta/2)+1}$.
The decrease in the number of conducting channels leads to suppression of the magnitude of $J_{max}$.

\begin{figure}[htbp]
\centering
\includegraphics[width=.4\columnwidth]{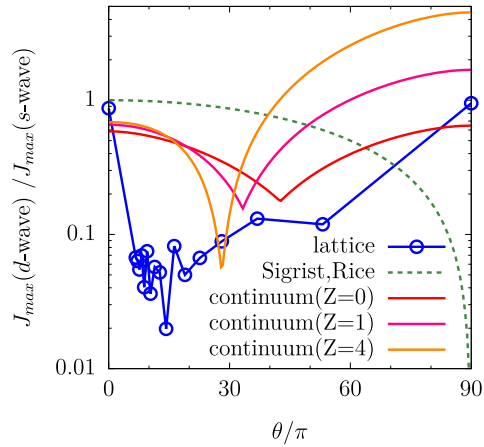}
\caption{The ratio of $J_{max}$ for a $d$-wave junction to that of an $s$-wave junction. The dotted line shows the result predicted by Sigrist and Rice's theory \cite{SigristRice}. We also plot the corresponding results obtained by the continuum model for $Z=0$, $Z=1$, and $Z=4$.}
\label{fig_normalized_Jc}
\end{figure}

\begin{figure}[htbp]
\centering
\includegraphics[width=.4\columnwidth]{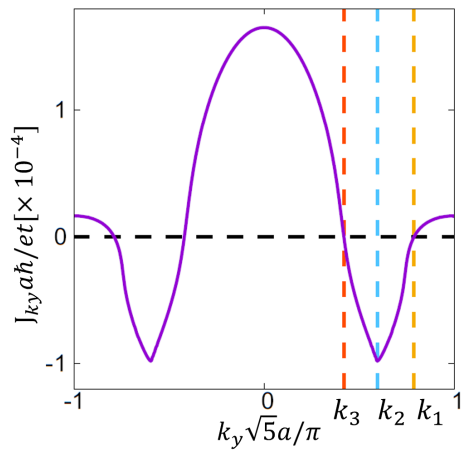}
\caption{Momentum-resolved Josephson current $J_{ky}$ for the (210) GB. 
We chose $\phi$ where $J$ gives the maximum Josephson current; $k_{1,2,3}$ are defined in figure \ref{fig_FermiSurface}.}
\label{fig_resolved_Josephson_current}
\end{figure}
\par
Next, we compare the $J_{max}$ of $s$-wave and $d$-wave pair potentials. 
The $J_{max}$ for the $d$-wave cases are found to be much smaller than those for the $s$-wave cases. 
To focus on the effect of the internal phase change, we plot $J_{max}$ of $d$-wave superconductor junctions normalized by that of $s$-wave junctions in figure \ref{fig_normalized_Jc}. 
We also show the $\theta$ dependence of the ratio of $J_{max}$ obtained by the phenomenological continuum model for comparison \cite{SigristRice}.
In the case of the phenomenological model, the suppression of $J_{max}$ originates from the cancellation of the 0-phase and $\pi$-phase current components where the sign of $J_1$ is positive and negative, respectively.
Similar cancellation occurs in the present lattice model.
\par
To confirm this calcellation, we show the $k_y$ dependence of the momentum-resolved Josephson current $J_{ky}$ in figure \ref{fig_resolved_Josephson_current}, where $k_y$ is the parallel momentum. 
We find that the sign of $J_{ky}$ changes at $k_y=k_1$ and $k_3$. 
These momenta correspond to the positions of the nodes.
That is, the sign change of the pair potential causes the sign change of the first-order Josephson coupling of $J_{ky}$. 
The cancellation of the first-order Josephson coupling then occurs in the integration by $k_y$, and the resulting Josephson current is suppressed.
Thus, the magnitude of $J_{max}$ for $d$-wave junctions is smaller than that for $s$-wave junctions.
\begin{figure}[htbp]
\centering
\includegraphics[width=.6\columnwidth]{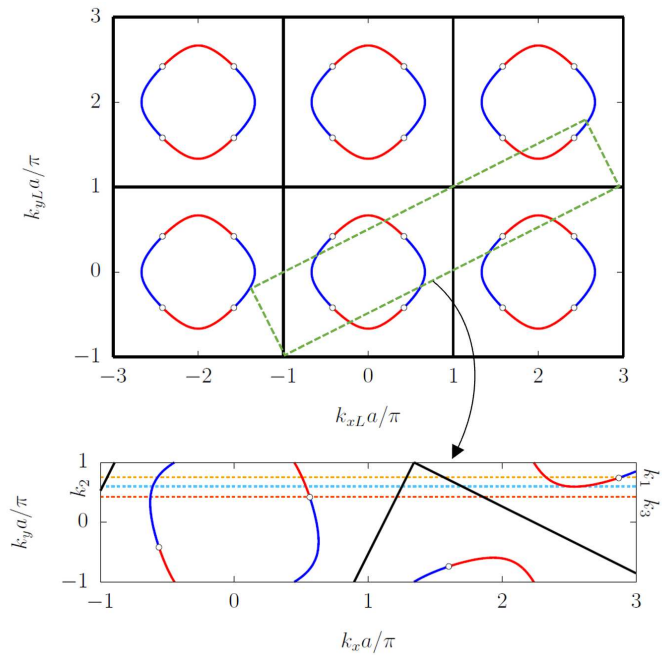}
\caption{Fermi surfaces and Brillouin zone for the (210) GB. The blue (red) lines correspond to the Fermi surfaces with positive (negative) pair potential; $k_1,k_2$ are the wavelengths at nodal points of the $d$-wave pair potential, and $k_2$ is the wavelength where the number of the Fermi surface changes.}
\label{fig_FermiSurface}
\end{figure}

\par
Although the reason for the sign change of the momentum-resolved current in the lattice model is the same as that in the continuum model, the $\theta$ dependence of the Josephson current shown in figure \ref{fig_normalized_Jc} exhibits different behaviors. 
For example, at lower $\theta$ values, the continuum model shows a single minimum value, whereas the lattice model shows oscillating behavior.
We find two reasons for this difference.
One reason is the folding of the Fermi surface (FS), as shown in figure \ref{fig_FermiSurface}.
This folding causes a change of the position of the nodes and leads to overlap of the FS.
The other reason is the change of the effective barrier potential. 
As $m$ increases, the number of conducting channels per unit length decreases.
The corresponding effective barrier potential then becomes pronounced.
The contribution to the 0-phase is then suppressed relative to the contribution of the $\phi$-phase, and the current is suppressed at lower $\theta$.
To observe these effects, we analyze the physical properties of the Josephson current in the lattice model in greater detail.
\begin{figure}[htbp]
\centering
\includegraphics[width=.4\columnwidth]{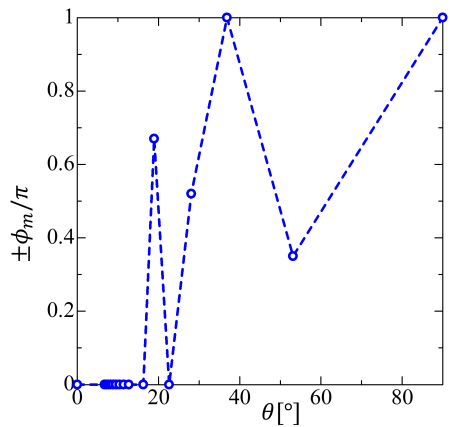}
\caption{Position of the free-energy minimum $\phi_m$ plotted as a function of tilting angle $\theta$.}
\label{fig_free_energy_minimum}
\end{figure}

First, we show the phase difference with the minimum free energy, which is obtained by integrating the CPR with respect to $\phi$, in figure \ref{fig_free_energy_minimum}.
A 0-junction appears in the lower-$\theta$ region, and a $\pi$-junction appears at higher-$\theta$ values.
We also observe $\phi$-junctions in certain places.
In the case of the continuum model, a $\phi$-junction only appears between the 0-junction at low $\theta$ and the $\pi$-junction at high $\theta$.
In the lattice model, a $\phi$-junction appears for the (410) GB at $\theta\simeq28^\circ$ between the 0-junction and $\pi$-junction regions.
However, a $\phi$-junction also appears in the cases of the (210) and (610) GBs.
This result is distinct from that of the continuum model.
\par

To clarify this point, we examine figure \ref{fig_resolved_Josephson_current} in greater detail.
In figure \ref{fig_resolved_Josephson_current}, a sign change of the momentum-resolved Josephson current occurs at the position of the nodes, as previously noted.
However, the position of nodes in the lattice model differs from that of the continuum model.
In the continuum model, a node appears at $k_y=k_F\sin(\pi/4\pm \theta/2)$ because the Fermi surface for a symmetric-tilt GB is just given by the rotation of the $k_x$- and $k_y$-axes.
However, in the lattice model, when $k_y=k_F\sin(\pi/4\pm \theta/2)$ is greater than the momentum of the BZ boundary, the FS is folded (figure \ref{fig_FermiSurface}).
In the case of the (210) GB, the nodal position is closer than in the case of the continuum model because the nodes at $k_y=k_1$ are on the folded FSs.
The area of the $\pi$-phase ($k_3<|k_y|<k_1$) then becomes small, and the negative contribution to the current decreases.
The $\phi$-phase then appears at $T=0.05T_c$, and the resultant Josephson current is smaller than that of the continuum model.
Similarly, a $\phi$-junction also appears in the case of the (610) GB between the 0-junction of the (510) and (710) GBs.
This effect is highly sensitive to the position of the nodal position in the BZ of the tilted structure; thus, it strongly depends on the chemical potential $\mu$. 
\par
A further difference between the lattice model and continuum model is the interference by the folded FSs.
Considering the (210) GB as an example, two FSs exist at $|k_y|>k_2$, as shown in figure \ref{fig_FermiSurface}, and the signs of the pair potentials on these two FSs are opposite.
Thus, when we consider the incident and transmitted quasiparticles in the (210) GB with $|k_y|>k_2$, the first order of the Josephson coupling is canceled because the quasiparticles' contributions to the current have opposite signs.
Actually, as shown in figure \ref{fig_resolved_Josephson_current}, the magnitude of $J_{ky}$ at $|k_y|>k_2$ is much smaller than that of $|k_y|<k_3$, where there is no folded FS.
\par
\begin{figure}[htbp]
\centering
\includegraphics[width=.4\columnwidth]{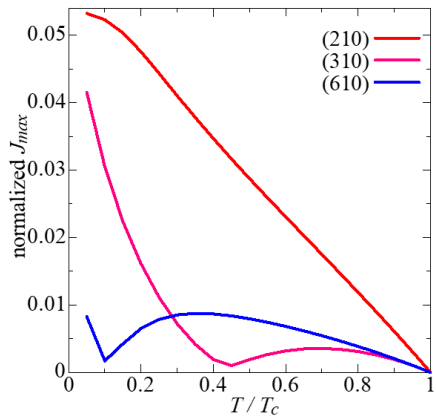}
\caption{Temperature dependence of $J_{max}$ for $d$-wave junctions normalized by $J_{max}(T=0.05T_c)$ for an $s$-wave. Tilting angle $\theta$ for the (210), (310), and (610) GBs is $\theta\simeq$ 53$^\circ$, 37$^\circ$, and 19 $^\circ$, respectively.}
\label{fig_Jc_temperature_dependence}
\end{figure}
\begin{figure}[htbp]
\centering
\includegraphics[width=.4\columnwidth]{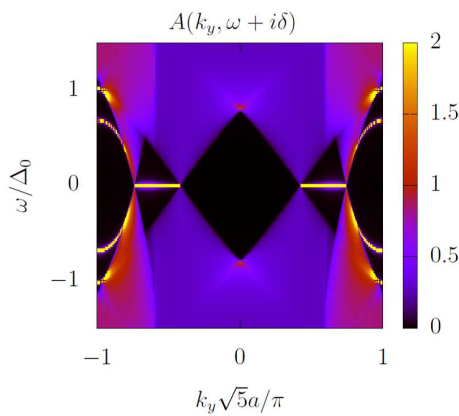}
\caption{Spectral function $A(k_y,\omega+i\delta)=-\frac{1}{\pi}\sum_{i=0,1,2}\mathrm{Im} G_{C,i,i}(k_y,\omega+i\delta)$ of the (210) GB. We chose $\phi=0, \delta=10^{-3}\Delta_0$.}
\label{fig_LDOS}
\end{figure}
Another feature of the proposed lattice model is that the normal conductance depends on its tilting angle.
In the present model, we do not introduce the barrier potential or the change of the hopping integral $t$ at the interface.
In the case of the continuum model without a barrier potential ($Z=0$), a 0--$\pi$ transition occurs at $\theta=45^\circ$; however, the $\pi$-phase appears at $\theta\simeq 37^\circ$ for the (310) GB.
Thus, an effective $Z$ exists in the present model because of the decreases of the conducting channel density.
We can also confirm this result from the temperature dependence of $J_{max}$ in figure \ref{fig_Jc_temperature_dependence}.
The temperature dependence of $J_{max}$ for the (210) GB exhibits saturation behavior at low temperatures and is similar to the previous result reported by Kulik-Omel'yanchuk \cite{Kulik}.
However, the temperature dependence of $J_{max}$ for the (310) and (610) GBs shows a rapid increase at low temperatures and nonmonotonic behaviors.
The rapid increase of $J_{max}$ is known to originate from the existence of the ZEABS at the interface.
The results also suggest that this ZEABS induces the $\pi$-phase.
As shown in figures \ref{fig_resolved_Josephson_current} and \ref{fig_LDOS}, the 0-phase and $\pi$-phase coexist in the momentum-resolved Josephson current and the area of the $\pi$-phase corresponds to that of the ZEABS \cite{TKJosephson,TKJosephson2}.
When the effective $Z$ increases with decreasing $\theta$, the contribution of the $\pi$-phase increases at lower temperatures.
The nonmonotonic temperature dependence due to the 0--$\pi$ transition and the rapid increase of the Josephson current at lower temperatures then appear.
\par
Finally, we note that a clear increase as decreasing $\theta$ in the $J_{max}$ at lower $\theta$ values is not observed in the lattice model.
As shown in figure \ref{fig_normalized_Jc}, the maximum Josephson current is not suppressed for the (100) GB ($\theta=0^\circ$) because it is a perfect crystal structure.
However, once the tilting angle becomes nonzero, the large suppression of the $J_{max}$ is observed.
In the present calculation, the Josephson current always passes through the sharing site at $x=0$ and its density decreases with decreasing $\theta$.
However, when $m$ becomes large, the edges of ($m10$) and ($\bar{m}10$) become nearly parallel and their interval is almost the same as the lattice constant $a$.
We then must consider the effect of direct hopping along the $x$ direction at the interface.
This hopping will make the (100) GB effective, likely resulting in an increase of the $J_{max}$.
\par
We here comment on the comparison between the present calculation results and the results of actual experiments. 
In many experimental studies of the GBs of HTSCs, the critical current has been reported to be severely suppressed with increasing tilting angle \cite{Held,Dimos,Hilgenkamp,Ivanov, Durrell,katase,Si,Iida2019}. 
The origin of this suppression cannot be explained using the previously reported continuum models, as shown in figure \ref{fig_normalized_Jc}.
By contrast, the suppression in the small $\theta$ region obtained in the present model suggests that the folding of the FS and the defect formation successfully reproduced the experimental trends. However, the recovery of $J_{max}$ with further increasing $\theta$ does not coincide with the experimental trends. 
This observation indicates that additional effects, such as diffusive scattering near the interface, occur in actual experiments corresponding to large-$\theta$ regions.
We believe the present model is important because it provides a guide for experiments showing that a mismatch at GBs strongly influences the Josephson current properties, even if the mismatch angle is slight.
\section{Summary}
In this paper, we calculated the Josephson current on the symmetric [001]-tilt GB of $d$-wave superconductor Josephson junctions with ($m10$)- and ($\bar{m}10$)-oriented surfaces on the lattice model.
By changing the tilting angle, we found a wide variety of CPRs, including $0$-, $\pi$-, and $\phi$-junctions. 
In addition to the suppression of the maximum Josephson current associated with the internal phase change of the pair potential, as observed in the continuum models, we found that further phase interference occurs because of the folding of the FS.
The obtained maximum Josephson current in the present lattice model is smaller than that in preexisting theories based on the continuum model in low-angle regions.
Because similar suppression of the critical current corresponding to maximum Josephson current at GBs has been reported in several experimental studies, the obtained results can serve as a guide to clarify the complicated transport mechanism in GBs.
\par
The roughness of the surface/interface is known to strongly influence the charge transport behavior in $d$-wave superconductor junctions \cite{yamada96,Golubov1998,Golubov1999}. 
The diffusive scattering near the interface destructively influences the contribution from ZEABSs because of the hidden odd-frequency odd-parity spin-singlet pairing \cite{Proximityd,Proximityd2,odd1,odd2,odd3}. 
Clarifying how the grain angle dependence of the critical current due to the interference of the phase from the folding of the BZ predicted in the present paper is influenced by the diffusive scattering would be an interesting topic for a future study. 
\par
In the present work, we have focused on the GB effect in $d$-wave superconductors.
Studies of iron-based superconductors in which $s_{\pm}$ pairing is a promising symmetry have also shown interesting results \cite{Mazin2008,Kuroki}. 
Because there are several theoretical works related to surface Andreev bound states, quasiparticle tunneling, and Josephson effects in $s_{\pm}$-wave superconductors \cite{Golubov2009,Onari2009,Ghaemi2009,NagaiHayashi,Linder2009,Ota2010,Huang2010,Bovkov,Seidel_2011,Burmistrova2013,Burmistrova_2013b,Burmistrova2015}, it is interesting to calculate the Josephson current in GB $s_{\pm}$ superconductor junctions. 

\begin{ack}
This work was supported by 
Scientific Research (A) (KAKENHIGrant No. JP20H00131),  
Scientific Research (B) (KAKENHIGrant No. JP20H01857), 
and
Scientific Research (Early-Career Scientists) (KAKENHIGrant No. 21K13854).
\end{ack}
\appendix
\section{Recursive Green's function method for ($m10$) surface}
In this appendix, we derive the Green's function in the junction of ($m10$) GB. 
To calculate this, we first calculate the surface Green's function at ($m10$) and ($\bar{m}10$) surfaces. 
For that purpose, we use the recursive Green's function method.
As given in the main text, the matrix elements for ($m10$) surface is given by
\begin{eqnarray}
\langle p,q|\mathcal{H}|p+1,q\rangle&=&
\begin{pmatrix}
0 & 0 &0&\cdots&0\\
t_x & 0&0 &\cdots&0\\
0 &t_x &0&\ddots&0\\
\vdots&\ddots&\ddots&\ddots&0\\
0&0&0&t_x&0
\end{pmatrix},\\
\langle p,q|\mathcal{H}|p,q\rangle&=&
\begin{pmatrix}
h_0 & t_y &0&\cdots&0\\
t_y^\dag & h_0&t_y &\cdots&0\\
0 &t_y^\dag &h_0&\ddots&0\\
\vdots&\ddots&\ddots&\ddots&t_y\\
0&0&0&t_y^\dag&h_0
\end{pmatrix},\\
\langle p,q|\mathcal{H}|p,q+1\rangle&=&
\begin{pmatrix}
0 & \cdots & 0 & t_x\\
0 & \ddots & 0 & 0\\
\vdots&\ddots&\ddots&\vdots\\
0&\cdots&0&0
\end{pmatrix},\\
\langle p,q|\mathcal{H}|p+1,q-1\rangle&=&
\begin{pmatrix}
0 & \cdots & \cdots & 0\\
\vdots & \ddots & \ddots & \vdots\\
0&0&\ddots&\vdots\\
t_y&0&\cdots&0
\end{pmatrix},
\end{eqnarray}
where
\begin{eqnarray}
h_0&=&
\begin{pmatrix}
-\mu & 0\\
0 & \mu
\end{pmatrix},\\
t_x&=&
\begin{pmatrix}
-t & \Delta(p)\\
\Delta^*(p) & t
\end{pmatrix},\\
t_y&=&
\begin{pmatrix}
-t & -\Delta(p)\\
-\Delta^*(p) & t
\end{pmatrix}.
\end{eqnarray}
By the Fourier transformation in $q$, we can diagonalize $\mathcal{H}$ in $q$,
\begin{eqnarray}
\langle p,k_y|\mathcal{H}|p+1,k_y\rangle&=&
\begin{pmatrix}
0 & 0 &0&\cdots&0\\
t_x & 0&0 &\cdots&0\\
0 &t_x &0&\ddots&0\\
\vdots&\ddots&\ddots&\ddots&0\\
t_ye^{iky}&0&0&t_x&0
\label{eq._Hamiltonina_p}
\end{pmatrix}\nonumber\\
&\equiv&T_x(k_y),\\
\langle p,k_y|\mathcal{H}|p,k_y\rangle&=&
\begin{pmatrix}
h_0 & t_y &0&\cdots&t_xe^{-ik_y}\\
t_y^\dag & h_0&t_y &\cdots&0\\
0 &t_y^\dag &h_0&\ddots&0\\
\vdots&\ddots&\ddots&\ddots&t_y\\
t_x^\dag e^{ik_y}&0&0&t_y^\dag&h_0
\end{pmatrix}\nonumber\\
&\equiv&H_0(k_y),
\end{eqnarray}
where $T_x$ and $H_0$ are $2(m+1)\times 2(m+1)$ matrices which include sublattice and particle-hole degree of freedom. 
\par
\begin{figure}[htbp]
\centering
\includegraphics[width=.7\columnwidth]{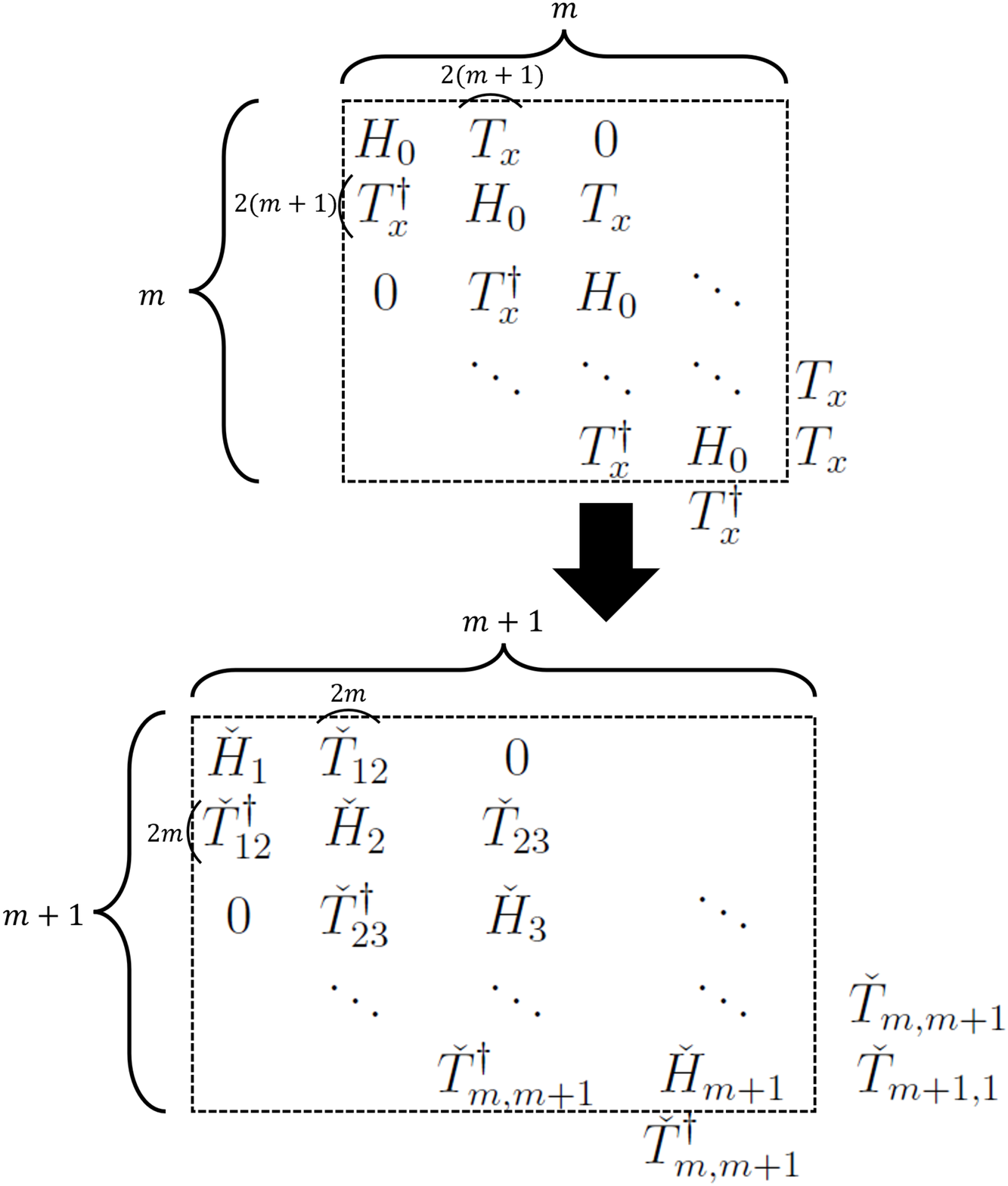}
\caption{$2(m+1)\times2(m+1)$ matrix $H_0$ and $T_0$ are represented by $2m\times2m$ matrix $\check{H}_{1,2,...,m+1},\check{T}_{12,23,...,(m+1,1)}$.}
\label{fig_matrix}
\end{figure}

Here, we suppose that the surface Green's function at $p=p_0+1$ in the $n$-layer system from $p=p_0+1$ to $p=p_0+n$ is known. 
Then, the surface Green's function at $p=p_0$ in the $(n+1)$-layer system from $p=p_0$ to $p=p_0+n$ is given by
\begin{align}
\label{eq._recursive_Green}
G^{(n+1)}_{s}(i\epsilon_n,k_y)=\left(i\epsilon_nI-H_0-T_xG^{(n)}_{s}(i\epsilon_n,k_y)T_x^\dag\right)^{-1},
\end{align}
where $G^{(n+1)}_{s}(i\epsilon_n,k_y)$ stands for the surface Green's function with Matsubara Frequency $\epsilon_n$ at $p=p_0$ ($p=p_0+1$) in the $(n+1)$-layer ($n$-layer) system.
Since it is trivial to obtain the Green's function of the 1-layer system by $G^{(1)}_{s}(i\epsilon_n,k_y)=(i\epsilon_nI-H_0)^{-1}$, we can calculate the surface Green's function in the system with any number of layers by the recursion relation in \eqref{eq._recursive_Green}. 
By repeating this recursive equation, surface Green's function converges as $G^{(n+1)}_{s}(i\epsilon_n,k_y)=G^{(n)}_{s}(i\epsilon_n,k_y)$. 
However, to get the convergence, we have to consider the system much larger than the coherence length. 
Thus, we can express the recursive equation in terms of a M\"{o}bius transformation. As we will explain later, 
$T_x$ must have the inverse matrix to do this. As seen from \eqref{eq._Hamiltonina_p}, $T_x$ is a lower triangular matrix without any diagonal components. 
Then, $T_x$ does not have the inverse matrix.
Thus, we express \eqref{eq._recursive_Green} in the another form. 
The matrix form of the total Hamiltonian is given by
\begin{eqnarray}
H&=&
\begin{pmatrix}
\ddots & \ddots &\ddots&&\\
\ddots & H_0& T_x &0&\\
\ddots &T_x^\dag & H_0 &T_x&\ddots\\
 &0 & T_x^\dag & H_0 &\ddots\\
& &\ddots&\ddots&\ddots\\
\end{pmatrix},
\end{eqnarray}
where $H_0$ and $T_x$ are $2(m+1)\times 2(m+1)$ matrices. Consider the $m$ contiguous layers of this system. 
Then, we can express this Hamiltonian by the $2m\times 2m$ matrices as shown in figure \ref{fig_matrix}.

By using these matrices, we make the recursion relation between surface Green's functions for the $n$-layer system and $(n+m)$-layer system,

\begin{align}
\check{G}_s^{n+m}=X_\bullet \check{G}_s^n,
\label{eq._check_Green}
\end{align}
where $\check{G}_s$ is the submatrix of $G_s$ including the sublattice $o=1$ to $m$ and therefore $\check{G}_s$ is $2m\times 2m$ matrix. Here, $_\bullet$ denotes the 
M\"{o}bius transformation defined by
\begin{align}
\begin{pmatrix}
a &b \\
c&d
\end{pmatrix}_\bullet z
\equiv (az+b)(cz+d)^{-1}.
\end{align}
$X$ is given by
\begin{align}
X=&\prod_{i=1}^{m+1}X_i,\\
X_i=&
\begin{pmatrix}
0& \check{T}_{m+i,i}^{-1}\\
\check{T}_{m+i,i}^\dag & (i\epsilon_n-\check{H}_i) \check{T}_{m+i,i}^{-1}
\end{pmatrix}
\end{align}
where the subscript of $\check{H}$ and $\check{T}$ is integers in modulo $m+1$.
These equations rewrite the process of adding $m$ layers containing $m+1$ sublattices into a process of adding $m$ sublattices $m+1$ times.
Since we can express the recursion relation in terms of a M\"{o}bius transformation in \eqref{eq._check_Green}, the problem to calculate the surface Green's function for semiinfinite layer is reduced into the eigenvalue problem of the matrix $X$.

$X$ is diagonalized as
\begin{align}
X = Q \begin{pmatrix}
\lambda_1 &&&{\huge O}\\
&\lambda_2 &&\\
&&\ddots&\\
{\huge O}&&&\lambda_{2m}\\ 
\end{pmatrix} Q^{-1} 
\end{align}
$\lambda_{1,2,\ldots,2m}$ are the eigenvalue of the matrix $X$. 
\begin{align}
    |\lambda_1|<|\lambda_2|<\ldots<|\lambda_{2m}|
\end{align}
\begin{align}
\check{G}_s^{\infty}=Q_\bullet I
\end{align}
We can calculate $G_s^{\infty}$ by adding a single layer from $\check{G}_s^{\infty}$.

\section*{References}
\bibliography{main}
\end{document}